# Look Who's Watching: Platform Labels & User Engagement on State-Backed Media Outlets

Samantha Bradshaw, Stanford Internet Observatory, Stanford University
Mona Elswah, Oxford Internet Institute, Oxford University
Antonella Perini, Oxford Internet Institute, Oxford University

**Abstract (275)** Recently, social media platforms have introduced several measures to counter misleading information. Among these measures are 'state media labels' which help users identify and evaluate the credibility of state-backed news. YouTube was the first platform to introduce labels that provide information about state-backed news channels. While previous work has examined the efficiency of information labels in controlled lab settings, few studies have examined how state media labels affect user perceptions of content from state-backed outlets. This paper proposes new methodological and theoretical approaches to investigate the effect of state media labels on user engagement with content. Drawing on a content analysis of 8,071 YouTube comments posted before and after the labelling of five state-funded channels (Al Jazeera English, CGTN, RT, TRT World, and Voice of America), this paper analyses the effect YouTube's labels had on users' engagement with state-backed media content. We found the labels had no impact on the amount of likes videos received before and after the policy introduction, except for RT which received less likes after it was labelled. However, for RT, comments left by users were associated with 30% decrease in the likelihood of observing a critical comment following the policy implementation, and a 70% decrease in likelihood of observing a critical comment about RT as a media source. While other state-funded broadcasters, like Al Jazeera English and VOA News, received fewer critical comments after YouTube introduced its policy; this relationship was associated with how political the video was, rather than the policy change. Our study contributes to the ongoing discussion on the efficacy of platform governance in relation to state-backed media, showing that audience preferences impact the effectiveness of labels.

***Key Words***: State Media, Platform Governance, Selective Exposure, YouTube, Labelling, Perceptions, and Credibility

**Corresponding Author:**
Samantha Bradshaw, Stanford Internet Observatory, Freeman Spogli Institute for International Studies, Encina Hall, Stanford University, Stanford, CA 94305-6055
Email: sam.r.bradshaw@gmail.com

## Introduction

The spread of disinformation online has become a critical question for political communication research and policymaking more broadly (Allcott and Gentzkow 2017; Freelon and Wells 2020; Ha, Andreu Perez, and Ray 2021; Lazer et al. 2018; Weeks and Gil de Zúñiga 2021). In order to combat its spread, social media platforms have introduced a range of policy responses (Taylor, Walsh, and Bradshaw 2018). Information labelling is one mechanism platforms have adopted to provide users with more information about the credibility of online news. In February 2018, YouTube was the first platform to implement a *state-funding* label for US audiences, providing notices below videos uploaded by news broadcasters that receive some level of government or public funding (Samek, 2018). In addition to state-funded labels, YouTube also provides *credibility* labels below scientific or politicized topics that have been subject to misinformation, linking to third party sources and additional fact-checked information (Mohan & Kyncl, 2018).

YouTube is not the only platform to apply *source* and *credibility* labels to news and information online. In 2016, Facebook introduced a credibility policy that labelled news stories containing false information with a "disputed" flag (Mosseri, 2017). This policy eventually replaced the "disputed" flag with a "related articles" notification, showing users fact checked information below stories that contained false information (Lyons, 2017). In June 2020, Facebook also implemented *source* labels for state-funded outlets that were "wholly or partially under the editorial control of their government" (Gleicher, 2020). Twitter began rolling out *credibility* labels in 2019 and 2020, including notices under tweets by politicians or world leaders that breach Twitter's rules (Twitter Safety, 2019), as well as notices under "synthetic and manipulated media" such as deep fake videos (Roth & Achuthan, 2020), and coronavirus mis-and-disinformation (Roth & Pickles, 2020). In August 2020, Twitter implemented a *source* labelling policy for state-affiliated media accounts from countries in the United Nations Security Council—Russia, China, the US, the UK, and France—however, in practice it only labelled Russian and Chinese outlets (Ha, 2020).

Labelling information is intended to make users critical readers of news by providing them with additional context about the information's source and accuracy. These labels aim to inoculate users by exposing them to weakened, refutational, or corrected arguments, which can help refute malicious or conspiratorial information (Cook, Lewandowsky, and Ecker 2017; Lim and Ki 2007; Linden et al. 2017; Nassetta and Gross 2020; Pfau et al. 1997). While there are only a few studies examining the impact of labels introduced by platforms, early evidence suggests mixed results. Warning labels can reduce the sharing of mis-and-disinformation (Mena, 2020), but if users have already been exposed to false claims, seeing a "disputed" flag after-the-fact will not alter their original belief (Pennycook, Cannon, & Rand, 2018). Labelling information can also lead to unintended consequences, such as the "implied truth effects," where unlabelled false claims are more likely to be interpreted as true (Pennycook et al., 2020). And emphasizing publisher information can increase the chance users view truthful claims as false (Dias, Pennycook, & Rand, 2020). While these studies shed light on the impact and effect of *correction* labels on user perceptions of news, few systematic studies have examined the effect of *source* labels, particularly in the context of state-backed media outlets. This is important given that state-backed media outlets—especially those from authoritarian regimes—have come under increasing pressure for spreading propaganda online (Diamond, Plattner, and Walker 2016; Elswah and Howard 2020; McFaul and Kass 2019; Molter 2020).

Studies related to selective exposure suggest people tend to consume news that aligns with their ideology, values, and pre-existing beliefs (Chaffee and Miyo 1983; Hart et al. 2009; Masip, Suau-Martínez, and Ruiz-Caballero 2018). Through this lens, one reason users select information that confirms their worldview is to avoid the cognitive dissonance of exposure to radically new or challenging information (Davis and Jurgenson 2014; Festinger 1957). Indeed, studies of selective exposure show that users are more likely to engage with social media content that is in line with their beliefs and opinions (Cinelli et al. 2020; Masip, Suau-Martínez, and Ruiz-Caballero 2018; Skoric, Zhu, and Lin 2018), and challenge, remove, or ignore conflicting information to reduce dissonance (Bai et al. 2019). Because social media labels can present ideas that conflict or challenge users' worldviews, we ask: when do state-funded labels affect users' perceptions and engagement with content from state-funded media outlets on YouTube?

The aim of this study is to understand the efficacy of state-media labels on users' perception of and engagement with state-backed media content online. First, we aim to understand under what conditions state-media labels inoculate users against potential manipulative or overt propaganda campaigns from state-backed media sources. Second, state-backed media are diverse in their content, coverage, and editorialization. However, when YouTube implemented its labelling policy, it received significant criticism for including editorially independent public broadcasters—like the BBC or PBS News—alongside state-backed propaganda outlets (Radsch 2018; Shaban 2018). Thus, we also aim to understand comparatively how state-media labels might affect the audiences of different state-backed channels.

Our findings contribute to the ongoing academic discussions about the efficacy of information labels on user perceptions by highlighting the importance of the audience. Our work compliments the growing body of empirical research that demonstrates the unintended consequences of labelling, strengthening the field's overall argument that labelling is not the ultimate solution to mis-and-disinformation. While previous experiments highlight the importance of timing or effect labels have on a user's ability to determine facts, by examining user-generated content we demonstrate that platforms should not expect homogenous effects of labels on their users, and that online users have varying ideologies and political leanings that should be considered when introducing misinformation countermeasures.

## Literature Review

### *Engagement, Exposure and Cognitive Dissonance*

The rich affordances of social media platforms allow users to interact with social media platforms in multiple ways, including liking or disliking, following, commenting, uploading, or sharing content online (Khan 2017). However, different motivations drive different types of interaction with online content (Kim and Yang 2017; Swani and Labrecque 2020).

Commenting on social media content is considered a higher act of public expression compared to watching, reading or liking content, which are considered lower level acts of engagement (Aldous, An, and Jansen 2019; Muntinga, Moorman, and Smit 2011). While engagement with social media content is considered a key indicator of the content popularity and the success of the content creator (Schreiner, Fischer, and Riedl 2019), it can also represent political information about users and audiences because of how users "selectively expose" themselves to content that is consistent with their worldview.

There is a rich corpus of literature examining how individuals play a role in exercising information preferences in both the online and offline world. In the offline world, research on selective exposure shows that people tend to select traditional media and broadcasting sources they want to consume, and choose to associate with certain political groups, including parties, community associations or candidates (Chaffee and Miyo 1983). The more an individual is interested in the subject the more likely they are to pay selective attention to sources and ideas that fit within their worldview (Berelson and Steiner 1964). Many of the studies on selective exposure are being updated to account for the affordances of social media platforms, showing that users are more likely to consume and interact with social media content that supports, rather than contradicts, their worldview (Cinelli et al. 2020; Masip, Suau-Martínez, and Ruiz-Caballero 2018; Quattrociocchi, Scala, and Sunstein 2016). In addition to exposing themselves to content selectively by following, watching or reading sources of news online, users are also more likely to *engage* with content that they perceive as believable and in line with their established beliefs (Kim and Dennis 2019). This includes commenting, liking, and sharing information that conforms to their political preferences.

One explanation for selective exposure is that users do not want to face the cognitive dissonance of exposure to new or challenging information (Chaffee and Miyo 1983; Cotton 1985; Cotton and Hieser 1980). Based on the Cognitive Dissonance Theory, people tend to selectively consume information that is in consistence with their cognition to avoid discomfort of holding two contradictory thoughts simultaneously (Festinger 1957; Jeong et al. 2019). The decision to read a certain online news story is motivated by the need for opinion reinforcement and an aversion to ideas that challenge one's perspectives (Garrett 2009). In the context of social media, users follow and share content that supports their beliefs, and eschew information that contradict their beliefs, known as *selective avoidance* (An, Quercia, and Crowcroft 2013; Garrett 2009; Johnson and Kaye 2013). Because of context collapse, users have expressed very real and jarring experiences when presented with unexpected information on social media (Davis and Jurgenson 2014). This negative experience motivates users to selectively consume certain information, while also selectively avoiding other online content and interactions in order to reduce the dissonance of conflicting information (Bradshaw and Howard 2018; Bai et al. 2019).

*Cognitive Dissonance, Belief Perseverance and News Corrections*

People avoid discomforting information in general. However, when people are confronted with additional contradictory evidence, they might adopt a dissonance reduction strategy, depending on the magnitude and importance of the discrepancy (Festinger 1957; Hardyck and Kardush 1968; Jost 2015; Vaidis and Bran 2018). Belief perseverance, or the idea that individuals maintain their beliefs despite new information that refutes its accuracy, is one way individuals attempt to avoid cognitive dissonance (Anderson, Lepper, and Ross 1980; Anglin 2019; Cohen, Aronson, and Steele 2000; Nestler 2010). In some studies, on belief perseverance, presenting individuals with additional facts does not always lead to a refutation, even they are based on a weak or inconclusive evidence (Anderson, Lepper, and Ross 1980).

When it comes to news corrections, many experiments have consistently shown corrections have little-to-no effect on a person's perception of truth (Thorson 2016). Studies have shown that even when presented with correct information, false beliefs continue to persist (Cobb, Nyhan, and Reifler 2013; Ecker et al. 2011; Thorson 2016). In some cases, exposure to corrections can even "backfire", with the belief in inaccurate information become even stronger after being presented with a correction (Nyhan and Reifler 2010). While concerns about the "backfire" effect continue, several studies have failed to find or replicate this effect (e.g., Nyhan et al. 2019; Schmid and Betsch 2019; Wood and Porter 2019). However, when subsequent backfires have been found, they have been mostly identified in small political or attitudinal subgroups (Swire-Thompson, DeGutis, & Lazer, 2020). Indeed, ideological cues have been shown to influence partisan perceptions of news credibility (Garrett 2009; Nyhan and Reifler 2010).

Nevertheless, much of the recent literature suggests audiences will accept authoritative cues over ideological ones and update their beliefs in the direction of the fact check (Bullock et al. 2013; Duncan 2020; Wood and Porter 2019). In the context of social media labels, some experimental research has demonstrated that being exposed to a "disputed flag" limits the likelihood of that user sharing information with a warning label (Mena 2020; Pennycook et al. 2020). Studies that focus on the effect of disputed flags have found a modest reduction in the belief of fake news stories (Gao et al. 2018; Pennycook et al. 2020). In particular, when disputed flags reduced the effect of mis-and-disinformation, users were exposed to the flag *before* the false or misleading information (Pennycook, Cannon, and Rand 2018). Similarly, Nassetta and Gross (2020) found that labels could mitigate the influence of misinformation if they were *noticed* by the users, demonstrating the importance of placement in addition to timing.

Despite the positive impact labels can have on inoculating users against mis-and-disinformation, scholars have also identified some unintended consequences of labelling practices. Unlabelled fake news stories were more likely to be perceived as true, a phenomenon called the "implied truth effect" (Pennycook et al. 2020). Others found that exposure to a "disputed" or "rated false" flag had no effect on the perceived accuracy of unlabelled false headlines but instead led to a decreased belief in the accuracy of true headlines (Clayton et al., 2019). When it came to *source* labels, increasing the visibility of publishers was "ineffective and perhaps even counterproductive", and showing participants headlines with publisher information had no significant effect on whether participants believed the information to be accurate (Dias, Pennycook, and Rand, 2020). Overall, research on the impact of news corrections and information labels presents inconclusive results, mainly from lab-controlled experiments. In this paper, we explore the effect of information labels from the users' point of view, evaluating user comments before and after the implementation of YouTube's state-funding labelling policy in 2018.

## Methodology

*Sampling Strategy*

Five state funded news channels were selected for this study: Al Jazeera English (AJE), China Global Television Network (CGTN), RT (formerly Russia Today), TRT World, and Voice of America (VOA) (see Table 1 and see the online supplemental file for more background information). We selected these channels for three reasons: first, despite being a transparency initiative, YouTube's labelling policy did not list which state-funded media outlets would be labelled. Several journalistic investigations at the time found that small or obscure state-funded channels were not accurately labelled (Kofman 2019), and YouTube also stated in its announcement that labelling would be an ongoing process (Samek, 2018). Thus, major state-funded news outlets were selected to ensure labels were accurately added to the videos uploaded by prominent state-funded channels.[i] Second, all five channels target US citizens with online content. Since YouTube's state-funded notice policy was not implemented globally and instead rolled out to an US audience starting on 2 February 2018, it was important to select channels that would have a large US audience who would notice the flag. Third, despite being state funded, all five channels are different in terms of their organizational culture, editorial policies, and practices, as well as geopolitical aims. Some state-funded news providers, like VOA and AJE, have criticized YouTube's notice policy for treating sources that provided independent journalism as comparable to known propaganda outlets like RT or Sputnik. By looking comparatively across a range of state-funded outlets, this paper can provide insight into whether these concerns are justified.

*Table 1:*
*A summary of the information available on the five channels*

| Channel | Launching Date | Country | Evidence on Government Interference | Funding Model | Social media Labels | Annual Budget in US$ |
|---|---|---|---|---|---|---|
| AJE | 2006 | Qatar | NA | State-funded | YouTube | NA |
| CGTN | 2000 | China | Available | State-funded | YouTube, Twitter, and Facebook | NA |
| RT | 2005 | Russia | Available | State-funded | YouTube, Twitter, and Facebook | 400 million |
| TRT World | 2015 | Turkey | Available | Public Broadcaster | YouTube | 77-155 million |
| VOA | 1942 | The US | NA | State-funded | YouTube | 218.5 million |

*Source*: Based on the available literature on these channels. An overview of these channels can be found in the supplemental file. *Note:* Government interference was identified based on the literature and articles. NA refers to the absence of available information related to this subject.

*Data Collection*

In this study, we focus on user generated YouTube comments to understand the efficacy of the state media labels. After selecting the five channels, we used a python script to query the YouTube API for a list of videos uploaded by the five outlets between 30 January 2018 and 1 February 2018, as well as 3 February 2018 and 5 February 2018. These dates represent three days before and three days after the notice implementation. There is no information on what time on 2 February 2018 YouTube implemented the notice policy. To ensure videos (and their subsequent comments) could be evaluated before and after the policy implementation, 2 February 2018 was excluded from the video and comment sample. In addition to the video titles and date of upload, we collected data about the video tags, number of views, number of video likes, number of video dislikes and number of comments. Table 2 summarizes the number of YouTube videos each channel uploaded during the study timeframe and provides summary statistics about engagement.

From this large sample of videos, we selected the top-three most commented on videos for each day in the timeframe. We selected a small sample of videos that were most widely discussed by users to ensure content was watched and consumed. In total, we selected 88 videos uploaded by the five state-funded media outlets on YouTube. These videos covered a range of political and economic news, including Trump's State of the Union Address, the downing of a Russian fighter pilot in Syria, child labour in China, and stock market forecasts.

*Table 2:*
*Videos, Views, (Dis)Likes & Comments (January 30-February 1 and February 3- February 5).*

| Channel/ State Funded Media | # of Videos Uploaded | # of Video Views | # of Video Likes | # of Video Dislikes | # of Comments |
|---|---|---|---|---|---|
| AJE | 52 | 1,403,493 | 11,220 | 13,587 | 5,977 |
| CGTN | 60 | 2,791,087 | 8,871 | 9,95 | 2,682 |
| RT | 57 | 2,471,910 | 24,460 | 6,256 | 18,150 |
| TRT World | 60 | 1,063,819 | 7,139 | 1,519 | 2,097 |
| VOA | 47 | 413,323 | 1,834 | 1,214 | 2,200 |

*Source*: Authors' analysis of videos collected through the YouTube API

Once we built a sample of videos, we queried the YouTube API for all the user comments associated with them. Because all the videos in the sample were still active videos at the time of data collection, users continued to watch and leave comments about this content. Since this paper is concerned with the impact the state-sponsored label had on users' comments and perceptions of these sources, we only selected comments left by users up to one month after the video was posted (until the end of February 2018). In total, 8,071 comments from 88 videos were coded for this study.

### *Coding & Typology Building: Assessing User Criticality*

In order to determine how users responded to content from state funded media organizations, we developed a typology using a grounded theory of coding (Charmaz, 2006). We developed a list of twelve discourse types that described the ways users actively engaged in discussion about content from the state-funded channels. High level sentiment scores were created to evaluate whether users (1) supported the channel or ideas presented in the video; (2) criticized the channel or ideas presented in the video; (3) neither supported nor criticized the channel or ideas presented in the video. The full codebook with descriptions can be found in the Appendix for this paper.

Three coders were trained on the typology through an interactive process where comments would be coded individually, decisions would be compared, disagreements would be discussed, and a final decision would be made. Coders individually coded 25% (or 2,021 comments) to measure intercoder reliability. Krippendorff's Alpha for high level code categories (supportive, critical, or neutral) was α = .87 suggesting substantial agreement between the coders. Krippendorff's Alpha for the subcategories was α = .81. This decrease in agreement was expected given the fuzzy nature of political communication and determining meaning from written comments. With an agreeable intercoder reliability established, the remaining 6,051 comments were split between the three coders who individually coded the remaining sample of comments. In addition to coding comments, the coders also coded videos they watched for level of politicization. This was done on an ordinal scale of not political, political, and highly political. Krippendorff's Alpha for the video scale was α = .88, suggesting strong agreement between coders. All disagreements for both videos and comments were discussed by the authors and a final decision was made.

## Analysis

### *State-Funded Flags & User Engagement*

First, to evaluate the impact of state-funded flags on user engagement with state-funded media channels, we collected data about the number of likes and dislikes videos received. We calculated the average number of likes and dislikes per video view, before and after the policy change (see Table 3). For all five channels, there was an overall decline in the number of likes videos received after the policy change, proportional to the views it received. We conducted a Welch Two Sample T-Test to examine the statistical significance between the difference in means between the average of likes and dislikes, before and after the policy implementation (see Table 3). While there was a decline in the number of likes received, the average in means before and after the policy change was only statistically significant for RT. There were no statistically significant relationships between the two averages of dislikes, before and after the policy implementation for any of the channels.

*Table 3:*
*Welch Two Sample T-Test. Comparison of likes and dislikes before and after YouTube state-funded policy implementation.*

|  |  | Full Dataset | Al Jazeera English | CGTN | RT | TRT World | VOA News |
|---|---|---|---|---|---|---|---|
| Likes (per 1000 views) | Before (n) | 12.4 | 10.4 | 1.8 | 18.4 | 16.6 | 5 |
|  | After (n) | 4.1 | 8.7 | 1.1 | 7.1 | 9.2 | 3.3 |
| Dislikes (per 1000 views) | Before (n) | 1.4 | 0.75 | 0.25 | 3.0 | 1.7 | 2.8 |
|  | After (n) | 0.7 | 0.14 | 0.14 | 0.09 | 1.1 | 2.1 |
| Likes | T-statistic (Pvalue) | -0.51 (0.60) | -0.09 (0.92) | -0.70 (0.49) | -1.99 (0.006)* | -0.43 (0.66) | -0.24 (0.80) |
| Dislikes | T-Statistic (Pvalue) | 0.73 (0.46) | -0.12 (0.90) | 1.63 (0.14) | -0.13 (0.89) | 0.89 (0.39) | -1.42 (0.17) |

*Note*: *indicates the coefficient estimate is statistically significant at the 99% confidence level.

## *State-Funded Flags & User Comments*

How did the introduction of the YouTube flag influence user comments and how critical they were of the content coming from these channels? We first compared the proportions of critical, supportive, and neutral comments before and after the policy change. Following the introduction of YouTube's state-funded labelling policy, user comments became less critical of content from AJE (34.6% to 25.4%) and RT (19.6% to 16.1%). At the same time, these channels showed an increase in the proportion of supportive comments after the policy change. AJE had the largest (26.8% to 41.9%) increase in the proportion of supportive comments after the policy change, while analysis of RT showed only a slight increase in supportive comments. In contrast, TRT World and CGTN received a larger proportion of critical comments after the implementation of YouTube's state-funded labelling policy. CGTN only saw a small (25.7% to 28.0%) increase in critical comments after the policy change, while TRT World had a very large (27.9% to 44.8%) increase in the proportion of critical comments. We conducted a Pearson's Chi-squared to assess the dependence between YouTube's policy change and the types of comments posted to videos and found a statistically significant relationship across the entire dataset (pvalue = 1.26e-07). The test statistic was also statistically significant when examining the dependence of these variables for AJE, RT, and TRT World (see Table 4).

*Table 4:*
*Proportion of comment sentiment before and after YouTube state-funded policy implementation, by state-funded channel*

|  |  | Full Dataset | AJE | CGTN | RT | TRT World | VOA News |
|---|---|---|---|---|---|---|---|
| Supportive Comments | Before (%) | 11.5% | 26.8% | 44.3% | 55.6% | 45.3% | 30.1% |
|  | After (%) | 36.1% | 41.9% | 39.8% | 56.1% | 23.5% | 33.7% |
| Critical Comments | Before (%) | 7.1% | 34.6% | 25.7% | 19.6% | 27.9% | 34.2% |
|  | After (%) | 15.7% | 25.4% | 28.0% | 16.1% | 44.8% | 34.7% |
| Neutral Comments | Before (%) | 8.1% | 38.9% | 30.0% | 24.8% | 26.7% | 35.1% |
|  | After (%) | 21.6% | 32.7% | 32.3% | 27.9% | 31.7% | 31.5% |
| Chi Squared | x-squared | 31.6 | 34.12 | 0.41 | 9.27 | 17.16 | 1.6 |
|  | Pvalue | 1.26e-07* | 3.91e-08* | 0.81 | 0.009* | 0.0002* | 0.4 |

*Notes*: * Indicates the coefficient estimate is statistically significant at the 99% confidence level3

We then performed a logit regression analysis asking: (1) does the YouTube policy change have an impact on the likelihood that a comment about content will be critical (see Table 5); and (2) does the YouTube policy change have an impact on the likelihood that a comment about the source will be critical (see Table 6)? We found that YouTube's policy change was associated with a lower likelihood of getting a critical comment about content by approximately 25% across all five state-funded channels. When we looked at the relationship for each news outlet, we found that a lower likelihood of receiving a critical comment was statistically significant for AJE (approximately a 35% decrease in likelihood) and RT (approximately a 30% decrease in likelihood). This relationship did not hold for all the channels. While CGTN, and VOA News showed no statistical significance, TRT World was associated with a statistically significant higher likelihood of getting a critical comment after the introduction of the YouTube flag—approximately a 100% increase in likelihood.

*Table 5*
*Logit Regression: what is the likelihood of getting a critical comment (general) after the policy implementation?*

|  | Full Sample | AJE | CGTN | RT | TRT World | VOA News |
|---|---|---|---|---|---|---|
| Intercept, coefficient (standard error) | -1.01* (0.05) | -0.63* (0.10) | -1.06* (0.27) | -1.41* (0.08) | -0.95* (0.24) | -0.65* (0.10) |
| After, coefficient standard error | -0.28* (0.06) | -0.44* (0.12) | 0.01 (0.32) | -0.25* (0.08) | 0.74* (0.26) | 0.08 (0.14) |
| After, Odds Ratio | 0.75 | 0.64 | 1.12 | 0.78 | 2.1 | 0.02 |
| Approximate (%) | -25% | -35% | +10% | -30% | +100% | +5% |
| Sample size (n) | 8,071 | 1584 | 230 | 4853 | 438 | 962 |

*Note:* * indicates the coefficient estimate is statistically significant at the 99% confidence level3

Additionally, we found that many of the comments in our dataset specifically criticized the channel itself. After the introduction of the label, many critical comments about the source mentioned the policy and discussed the label in the context of government propaganda. We measured and compared the proportion of these critical source comments before and after the implementation of the policy and found that RT was associated with a 70% decrease in likelihood of observing a critical comment about RT as a media source following the introduction of the label (see Table 6).

*Table 6*
*Logit Regression: what is the likelihood of getting a critical comment (source) after the policy implementation?*

|  | Full Sample | AJE | CGTN | RT | TRT World | VOA News |
|---|---|---|---|---|---|---|
| Intercept, coefficient (standard error) | -2.60* (0.08) | -2.82* (0.20) | -2.37* (0.43) | -2.27* (0.11) | -2.60* (0.42) | -3.71* (0.31) |
| After, coefficient standard error) | -0.46* (0.10) | -0.14 (0.24) | -0.25 (0.52) | -1.14* (0.14) | 1.10* (0.44) | 0.15 (0.41) |
| After, Odds Ratio Approximate (%) | 0.63 -35% | 0.87 -10% | 0.78 -20% | 0.31* -70% | 3.0 +200% | 1.16 +15% |
| Sample size (n) | 8,071 | 1584 | 230 | 4853 | 438 | 962 |

*Note:* * indicates the coefficient estimate is statistically significant at the 99% confidence level3

Next, we controlled for other variables that might have impacted the likelihood of receiving a critical comment. Highly politicized videos, such as those on war and conflict, immigration policy, or the Trump administration, might generate more critical comments compared to topics like sports, lifestyle news, or entertainment. Thus, we performed a third logit regression (see Table 7) asking: does the YouTube policy change have an impact on the likelihood that a comment will be critical after controlling for how political a video is? This analysis allowed us to control for the possibility that political videos were not randomly distributed before and after the event. In the first regression, YouTube's policy change was associated with approximately a 25% lower likelihood of a critical comment. After controlling for how political the video was, the likelihood of receiving a critical comment after the implementation of YouTube's policy remained unchanged. However, AJE and VOA News were associated with a statistically significant likelihood of getting a critical comment if the video was classified as highly political. For AJE, there was approximately a 20% decrease in likelihood of observing a critical comment if the video was highly political. For VOA News there was an approximately 70% decrease in the likelihood of observing a critical comment. These findings will be discussed in greater detail the next section of the paper.

*Table 7*
*Linear Regression: What is the likelihood of getting a critical comment when controlling for the politicization of video content?*

|  | Full Sample | AJE | CGTN | RT | TRT World | VOA News |
|---|---|---|---|---|---|---|
| Intercept, coefficient | -0.96* | -1.03* | -1.03* | -11.33 | -2.87* | 0.48 |
| (standard error) | (0.12) | (0.16) | (0.34) | (196.97) | (1.08) | (0.44) |
| After, coefficient | -0.32* | -0.48* | -0.11 | -0.23* | 0.72* | -0.04 |
| (standard error) | (0.06) | (0.12) | (0.33) | (0.09) | (0.27) | (0.14) |
| After, Odds Ratio | 0.72 | 0.62 | 1.11 | 0.79 | 2.06 | 0.90 |
| Approximate % | -25% | -40% | +11% | -20% | +105% | -10% |
| Highly Political, coefficient | 0.03* | -0.79* | -0.33 | 9.90 | 2.06 | -1.14* |
| (standard error) | (0.07) | (0.16) | (0.40) | (196.97) | (1.06) | (0.43) |
| Highly Political, Odds Ratio | 0.75 | 0.81 | 0.71 | 2001.88 | 7.84 | 0.32 |
| Sample size (n) | 8,071 | 1584 | 203 | 4853 | 438 | 962 |

*Note:* *indicates the coefficient estimate is statistically significant at the 95% confidence level3

## Discussion

How did YouTube's state-funding labels affect user engagement with content from these channels? And how did YouTube's state-funding label affect user comments about these channels and the content they produce? Results varied across our channels, which could be explained by the editorialization of these channels that attract different segments of YouTube users. First, the labels had no impact on the amount of engagement videos received before and after the policy introduction, except for RT, where there was a statistically significant decrease in the average of likes videos received after the policy introduction. This is worth noting in relation to the analysis on comments by users, where we found a 30% decrease in the likelihood of observing a critical comment after the policy change and a 70% decrease in the likelihood of observing a critical comment about the source specifically. On the one hand, RT received fewer likes after the policy change, but on the other hand, users also became less critical of the content in their comments. While it seems like these two findings do not comport, we suggest they actually capture two different effects of the policy change on two different audience segments of RT.

Research on RT has demonstrated that the channel targets audiences who share anti-Western and anti-establishment ideologies (Elswah and Howard 2020). At the same time, RT is the most popular state-funded news channel, with the most YouTube engagement especially compared to the other four channels in the study. RT also spends a significant amount of funds growing audiences and disseminating news content on social media specifically targeted towards American audiences (Bradshaw, DiResta and Miller, forthcoming). Thus, people who watch RT might not only belong to an ideologically similar subgroup of users but might stumble upon its content through YouTube's recommendation algorithm or click-throughs from other social media websites like Twitter or Facebook. Users who do not belong to RT's dedicated audience base might be less invested in the content and less likely to comment or engage in discussion about its coverage. However, they might engage in smaller actions, such as liking (or disliking) the video. Thus, the flag might have discouraged broader viewers from liking RT content when the state-funding label was presented. For RT's dedicated audience base who are more likely to comment on videos, the introduction of YouTube's state-funding flag could have challenged their worldview. To avoid the cognitive dissonance of the label, belief perseverance could explain the drop in critical comments about the channel and its content following the introduction of the labels.

In contrast to RT, TRT World was an outlier in the dataset, where the likelihood of seeing a critical comment increased by 100% following the introduction of YouTube's policy change. Unlike RT, TRT World is not as widely recognized as being under the editorial control of the Turkish government (Stalinsky, 2019). Because TRT World is also, by comparison, a relatively new channel, its audience base might not be as established as that of a channel like RT which has spent years tailoring content to a particular kind of audience. Thus, the introduction of YouTube's policy might have made the average YouTube user more critical of the content coming from TRT World.

Finally, across the entire dataset, the analysis showed there was a 25% decrease in the likelihood of observing a critical comment after the policy change. After controlling for how political videos were this relationship did not hold for AJE and VOA News, where both channels were associated with a statistically significant decrease in critical comments if the video was classified as highly political. This is a notable finding for these two channels that have greater levels of independence and adherence to professional standards of journalism, because it suggests that users do not necessarily care about the label, but instead care about the kinds of content and coverage these channels provide. Thus, concerns that the state-funded labels would decrease user trust in public broadcasters or other state-funded channels that have greater levels of editorial independence are not supported by our findings.

## Study Limitations & Future Research

There are several important limitations to this research that should moderate the findings. First, it is important to note that YouTube's state-funded labelling policy is not global. The policy was first introduced in the United States on 2 February 2018 but has subsequently been rolled out in 21 other countries. Given that there is no way to firmly connect a YouTube user to a geographic location, it is not possible to know that all of the comments in the dataset came from people who saw the flag. While channels were selected because of their focus on US politics and global reach, not all of the comments in the dataset would have come from individuals who saw the flag. Notices could also have a different impact on different audiences, and further research should be done to evaluate the impact of state-funded flags in non-western contexts.

Second, when it came to sampling channels, we selected prominent channels that could be verified as having the flag accurately applied. However, this has meant that other smaller channels or subchannels were left out of the study. For smaller or more obscure subchannels funded by state-organizations, users might be less aware of how the content they consume is being funded, and the policy might have had a different effect on the audiences of these channels.

Third, there are many factors that lead individuals to agree or disagree with news and information online. Comments and engagement metrics can provide some insight into how people consume political information, but unlike conducting an experiment, we do not have as much information about the audience makeup including demographics or pre-existing political beliefs. By examining user-generated content, we can observe genuine user interactions on social media platforms. Here, we have found different effects of labelling across the audiences of state-backed media outlets. The findings of our research can help inform the direction of future experimentation and research, where studies examining audience demographics and political values can enrich our understanding and provide nuance to the efficacy of information labels in the context of state-backed media.

## Conclusion

While labelling information seems like a sensible strategy to help users navigate an increasingly complex digital information ecosystem to inoculate users against attempts of persuasive attacks, these policies do not always work as intended. Rather, they might challenge the cognitive comfort of many online users who have selectively chosen exposure to state-backed media channels because of their editorialization. Research on the impact of labelling information on social media has shown both mixed and unintended effects: where information labels have shown a modest reduction in user's sharing and believing false information (Mena, 2020; Pennycook, Cannon, & Rand, 2018), other studies have found unlabelled information, even when false, is more trusted by users (Pennycook et al., 2020). Also, seeing labels reduces people's overall trust in factually correct and professional journalism (Clayton et al., 2019). While it is beyond the scope of this study to experimentally test the effect of these labels on users' cognition, the comments on RT have shown many examples of belief perseverance among some users who either trivialized the labels or bolstered support for the channel's content. Given the growing efforts by social media platforms to help users navigate an increasingly complex digital information ecosystem, understanding the effect of these labels is critical to improving platform initiatives to combat the spread of propaganda and disinformation online.

This study contributes to the growing body of empirical work examining the effects of labelling information on user perceptions and behaviors. By analyzing YouTube comments across five state-funded channels (AJE, CGTN, RT, TRT World, VOA) before and after the policy change, we identified some mixed results and unintended consequence. First, different audiences reacted differently to the state-funded label. For RT, while there was a decrease in the number of likes videos received after the implementation of YouTube's policy, comments on RT became simultaneously less critical. For RT's dedicated audience base—a specific subgroup of users who already do not trust mainstream media—there appeared to be belief perseverance among them. These users reduced the dissonant cognitions caused by the labels by devaluating and decreasing its importance. For TRT World, which is not as readily recognized as a propaganda outlet, the introduction of the label increased the likelihood of observing a critical comment. Thus, while YouTube's policy might have contributed to some users being more critical of content from known propaganda outlets, it might have simultaneously led to trivialization by specific audience subgroups.

Second, news outlets, politicians and activists criticized YouTube's state-funded policy for treating independent public broadcasters the same as known propaganda outlets like CGTN, RT, and TRT World. However, for AJE and VOA News, broadcasters who have a greater level of independence, the likelihood of receiving a critical comment after YouTube's policy was associated with how political the video was, rather than the policy change. This finding aligns with other experimental research that found providing information about publishers did not affect user's attitudes or beliefs about news (Dias, Pennycook, & Rand, 2020). Similarly, YouTube's policy treats different audiences in the same way. As shown by our findings, this can have mixed consequences. For audiences whose cognitions or viewpoints are challenged, labels might trigger cognitive dissonance and strategies for its reduction.

In conclusion, YouTube's state-funding labelling practices have had mixed results. Despite various studies highlighting unintended consequences of labelling practices, platforms continue to implement these policies. This is problematic for two reasons. Firstly, one reason labelling is an attractive solution is because it puts the onus on the user to decide what information is good or bad and absolves social media platforms of editorial responsibility. Given their power to shape and influence how people consume and internalize news and information, platforms must be more transparent and ensure greater levels of oversight for their actions and decisions. Secondly, in order to actually address the spread of propaganda and disinformation online, private self-regulation strategies should be tested and empirically supported. Of course, scientific inquiry takes time and the challenges posed by platforms for our news and information ecosystem need to be imminently addressed. But this limitation could be overcome if the platforms make better efforts to work with scholars and publish the results of any internal testing they have done.

Platform Labels & User Perceptions                    Preprint Version – Paper Accepted at ABS

[i] We verified that all five state-funded outlets received a label by reading through tech news coverage about the policy change and exchanged emails with an US journalist investigating the accuracy of YouTube's policy implementation. Correspondence with a ProPublica reporter who provided a list of state-funded channels that did not have the YouTube label added.